\documentclass[11pt,a4paper]{article}
\pdfoutput=1
\usepackage{jcappub}
\usepackage{fancyhdr}
\usepackage{wrapfig}
\usepackage{amsmath,amsthm,amssymb}

\makeatletter
\def\@fpheader{\relax}
\makeatother

\usepackage{enumerate}
\usepackage{color}
\usepackage{graphicx}
\usepackage{caption}
\usepackage{subcaption}
\usepackage{sidecap}

\def\beq{\begin{equation}}
\def\eeq{\end{equation}}
\def\bea{\begin{eqnarray}}
\def\eea{\end{eqnarray}}

\def\bwt{\begin{widetext}}
\def\ewt{\end{widetext}}

\begin{document}

\title{Off-shell Dark Matter: \\ {\it A Cosmological relic of Quantum Gravity}}

\author[a,b]{Mehdi Saravani}
\author[a,b]{, Niayesh Afshordi}
\affiliation[a]{Perimeter Institute for Theoretical Physics, 31 Caroline St. N., Waterloo, ON, N2L 2Y5, Canada}
\affiliation[b]{Department of Physics and Astronomy, University of Waterloo, Waterloo, ON, N2L 3G1, Canada}
\emailAdd{msaravani@pitp.ca}
\emailAdd{nafshordi@pitp.ca}

\abstract{

We study a  novel proposal for the origin of cosmological cold dark matter (CDM) which is rooted in the quantum nature of spacetime. In this model, off-shell modes of quantum fields can exist in asymptotic states as a result of spacetime nonlocality (expected in generic theories of quantum gravity), and play the role of CDM, which we dub off-shell dark matter (O$f$DM). However, their rate of production is suppressed by the scale of non-locality (e.g. Planck length). As a result, we show that O$f$DM is only produced in the first moments of big bang, and then effectively decouples (except through its gravitational interactions). We examine the observational predictions of this model: In the context of cosmic inflation, we show that this proposal relates the reheating temperature to the inflaton mass, which narrows down the uncertainty in the number of e-foldings of specific inflationary scenarios. We also demonstrate that O$f$DM is indeed cold, and discuss potentially observable signatures on small scale matter power spectrum. 
}
\maketitle

\section{Introduction} \label{introduction}

A vast range of observations in Astrophysics and Cosmology have now provided concrete evidence for the existence of cosmological cold dark matter (CDM), which appears to make up the majority of mass density in our universe (only second to the mysterious dark energy). Rotation curves of galaxies (e.g. \cite{Persic:1995ru}), gravitational lensing (e.g. \cite{Clowe:2006eq}), and Cosmic Microwave Background (CMB) \cite{Ade:2015xua,Hinshaw:2012aka} all indicate that General Relativity with ordinary (or known) matter is not consistent with observations. It is worth noting that, unlike dark energy, evidence for the existence of CDM ranges from cosmological to galactic (i.e. six orders of magnitude) in physical scale. 

Since all the observational evidence for CDM is through its gravitational interactions, it has been tempting to explore a modification of Einstein gravity as a substitute (e.g. \cite{Milgrom:1983ca,Bekenstein:2004ne,Moffat:2004bm,Moffat:2005si}). However, given the range of observational data matched by CDM (in particular, the precision measurements of CMB anisotropy power spectrum \cite{Ade:2015xua,Hinshaw:2012aka}) it has become nearly impossible to fit the data with any modified gravity alternative (which does not have an effective built-in dark matter component) \cite{Clifton:2011jh}.

As a result, the most popular approach has been to consider CDM as a new (beyond Standard Model) weakly interacting particle. There is strong evidence that CDM particle has to be (at most) weakly interacting with the Standard Model, as otherwise it should have been detected by now, through various astrophysical or terrestrial probes (see, e.g. \cite{Feng:2010gw}). It also has to be sufficiently cold, as there is no evidence for a thermal cut-off in the cosmological matter power spectrum, down to sub-Mpc scales \cite{Viel:2013fqw}. It is quite remarkable that a simple assumption of adding a non-relativistic (and non-interacting) dark matter is compatible with all the cosmological observations.

Here, we study a rather different approach, first proposed in \cite{Saravani:2015rva}, which we shall refer to as off-shell dark matter (O$f$DM) in this paper. In this proposal, CDM originates from considering quantum gravitational effects on the evolution of quantum fields. These effects manifest themselves through modifying the evolution law of quantum fields to a non-local evolution described by a causal non-local operator $\widetilde \Box$ which substitutes the role of D'alembertian. 

Let us outline some features of this model. First, this non-local modification results in the appearance of a new set of modes (or excitations) associated to each field. In fact, modification of a field with mass $M$ leads to two sets of modes:
\begin{enumerate}
\item Modes with mass $M$, called on-shell.
\item A continuum of massive modes with mass higher than $M$, called off-shell.
\end{enumerate}
We call the original mass of the field ($M$) ``intrinsic mass''. In other words, intrinsic mass is the mass of the on-shell modes (or the least value mass of the excitations). 

The important property that differentiates these two sets of modes and points to the direction of dark matter is the following: {\it transition rate of any scattering including one (or more) off-shell mode(s) in the initial state is zero}. 
This property makes off-shell modes a natural candidate for CDM, simply because they cannot be detected through non-gravitational scattering experiments \cite{Saravani:2015rva}. In fact, they can be produced by scattering of ``on-shell" particles, but they do not scatter, annihilate or decay. As such, the only way to detect these particles is through their gravitational signatures.

In the next section, we will review the important features of this model. Section \ref{Dark Matter Production} is dedicated to the production of O$f$DM in the context of inflation and reheating. We will discuss the effect of O$f$DM on matter power spectrum in Section \ref{Cold OFDM}. Finally, Section \ref{conclusion} concludes the paper.

\section{Review of O$f$DM}
Let us start this section by the following question: If off-shell modes of matter can be produced by the scattering of on-shell modes, while the reverse does not happen, shouldn't we see any signature of this in scattering experiments, for example in Large Hadron Collider (LHC)?
\begin{figure}
\begin{center}
\includegraphics[width=0.95\linewidth]{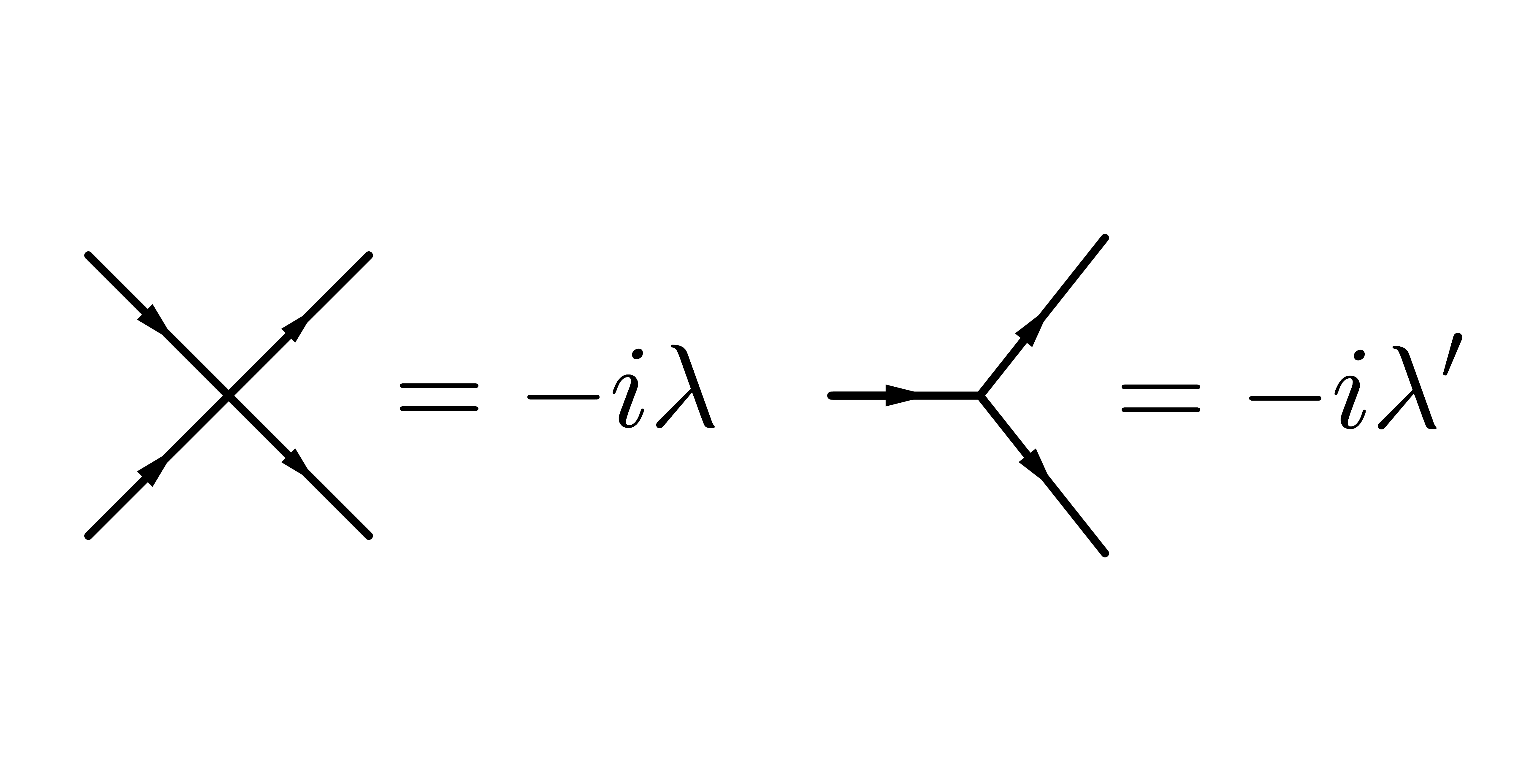}
\end{center}
\caption{A simple annihilation process (on left) and decay process (on right).}
\label{scattering}
\end{figure}
In other words, whenever we perform scattering experiments, a part of the incoming energy must transfer to off-shell modes and become undetectable. Shouldn't we have already seen this effect by now?

In order to answer this question, consider a simple annihilation or decay process (Figure \ref{scattering}). First, let us define the following quantities: 
$\sigma_{1F}$ ($\Gamma_{1F})$ is the cross-section (rate) of producing one off-shell particle and one on-shell particle and
$\sigma_{O}$ ($\Gamma_{O})$ is the cross-section (rate) of producing purely on-shell particles. If we assume that the energy of the process is much higher than the intrinsic mass of the out states, $E_{\rm CM} \gg M$ (as we will see later, this is the relevant regime for dark matter production), following the results in \cite{Saravani:2015rva}, we arrive at\footnote{$\delta_+(p^2)\equiv \delta(p^2)\theta(p^0)$}
\beq
\label{production_rate}
\frac{\Gamma_{1F}}{\Gamma_O}=\frac{\sigma_{1F}}{\sigma_O}
=\frac{\int d^4p_1d^4p_2 2\pi \delta_+(p_1^2)\widetilde W(p_2)\delta^4(q-p_1-p_2)}{\int d^4p_1d^4p_2 2\pi \delta_+(p_1^2)2\pi\delta_+(p_2^2)\delta^4(q-p_1-p_2)}
\eeq
where $q$ is the incoming energy-momentum and
 $\widetilde W(p)$ is given in terms of the spectrum of non-local operator $\widetilde \Box$ 
 \bea
 \widetilde W(p)=\frac{2{\rm Im}~ B(p)}{|B(p)|^2}\theta(p^0),\\
\widetilde \Box e^{ip\cdot x}=B(p)e^{ip\cdot x}.
 \eea
Note that $\widetilde W(p)$ is the two point correlation function (or Wightman function) of the field in the momentum space (see Section 4 in \cite{Saravani:2015rva})
\beq
\langle0|\hat \psi(x)\hat \psi(y)|0\rangle=\int \frac{d^4p}{(2\pi)^4}\widetilde W(p)e^{ip\cdot (x-y)}
\eeq


Equation \eqref{production_rate} can be simplified further if we assume that the energy scale of the scattering\footnote{Throughout this paper we are using $(-+++)$ signature for the metric.} $-q^2\equiv E^2_{CM}$ is much lower than the non-locality scale $\Lambda$ defined through $\widetilde \Box$. In this regime,
\bea
B(q)&=&-q^2+\mathcal{O}\left(\frac{q^4}{\Lambda^2}\right)\\
{\rm Im}~B(q)&=&a\frac{q^4}{\Lambda^2}+\mathcal{O}\left(\frac{q^6}{\Lambda^4}\right).
\eea
For $a\neq 0~$\footnote{Another possibility would be that $a=0$. In that case, the leading term to the imaginary part of $B$ comes in $6^{th}$ order. We will not pursue this possibility in this paper.}, $\Lambda$ can be redefined to set $a=\frac{1}{2}$.

With this assumption, we can make use of the Taylor expansion of $\widetilde W$
\beq\label{wtilde}
\widetilde W(q)=\frac{1}{\Lambda^2}+\mathcal{O}\left(\frac{q^2}{\Lambda^4}\right), ~~M^2\ll-q^2\ll\Lambda^2,
\eeq
to finally get (to the leading order)
\beq
\label{production_rate_simplified}
\frac{\Gamma_{1F}}{\Gamma_O}=\frac{\sigma_{1F}}{\sigma_O}= \frac{1}{4\pi}\left(\frac{E_{\rm CM}}{\Lambda}\right)^2,
\eeq
where $E_{\rm CM} \ll \Lambda$ is the centre of mass energy of the incoming particle(s). Note that for a decay process, $E_{\rm CM}$ is the mass of the decaying particle. Although, we derived \eqref{production_rate_simplified} for simple interactions of Figure \ref{scattering}, it is generally correct (up to order one corrections) as long as $E_{\rm CM}$ is much higher than the intrinsic mass of the intermediate particle(s) in Feynman diagrams. 

Now, let us define $\sigma_{2F}~(\Gamma_{2F})$ to be the cross section (rate) of producing two off-shell particles in the out state (Figure \ref{scattering}). Then,

\bea
\label{production_rate_2}
\frac{\Gamma_{2F}}{\Gamma_O}=\frac{\sigma_{2F}}{\sigma_O}&&=\frac{\int d^4p_1d^4p_2 \widetilde W(p_1)\widetilde W(p_2)\delta^4(q-p_1-p_2)}{\int d^4p_1d^4p_2 2\pi \delta_+(p_1^2)2\pi\delta_+(p_2^2)\delta^4(q-p_1-p_2)}\notag\\
&&=\frac{1}{48\pi^2}\left(\frac{E_{CM}}{\Lambda}\right)^4
\eea

As we see, adding one more off-shell particle in the final state suppresses the cross section by another factor of $\left(\frac{E_{CM}}{\Lambda}\right)^{2}$. So, the rate of two off-shell particles production is suppressed by a factor of $\left(\frac{E_{CM}}{\Lambda}\right)^{2}$ compared to one off-shell particle production. 


Before going any further, let us discuss the typical mass of the off-shell particle produced in Figure \ref{scattering}. For one off-shell particle production, the mass distribution of the produced off-shell particle is given by 
\beq
P_{1F}(m)=N\int d^4p_1d^4p_2\delta_+(p_1^2)\widetilde W(p_2)\delta^{(4)}(q-p_1-p_2)~m \delta(p_2^2+m^2),
\eeq
Where $N$ is the normalization factor. Using \eqref{wtilde} it reduces to
\beq\label{mass_density}
P_{1F}(m)=\frac{4m}{E^2_{\rm CM}}\left(1-\frac{m^2}{E^2_{\rm CM}}\right) ~~0<m<E_{\rm CM},
\eeq
assuming that the off-shell particle is intrinsically massless (or that its mass is much smaller than $E_{CM}$). For production of two off-shell particles, the mass distribution is given by
\beq
P_{2F}(m)=N'\int d^4p_1d^4p_2\widetilde W(p_1)\widetilde W(p_2)\delta^{(4)}(q-p_1-p_2)~ m \delta_(p_2^2+m^2),
\eeq
which reduces to 
\beq
P_{2F}(m)=\frac{48m}{E_{CM}^2}\bigg[\frac{1}{4}-\frac{1}{4}\left(\frac{m}{E_{CM}}\right)^4-\left(\frac{m}{E_{CM}}\right)^2\sinh^{-1}\left(\frac{E_{CM}^2-m^2}{2mE_{CM}}\right)\bigg].
\eeq
In both cases, the typical mass of the produced off-shell particles is $\sim E_{CM}/2$.

Now, we can estimate how likely it is to produce off-shell particles in LHC experiments. If we set $\Lambda \sim M_{\rm P}\equiv \frac{1}{\sqrt{8\pi G}} \sim 10^{18} ~{\rm GeV}$ and $E_{\rm CM} \sim 1~{\rm TeV}$ (LHC energy scale), we realize that the rate of producing off-shell particles in LHC is $10^{-31}$ lower than the rate of a normal scattering happening. In other words, out of $10^{31}$ scatterings in LHC, on average one results into the production of an undetectable particle (off-shell mode), explaining why O$f$DM could be well-hidden from high energy physics experiments.

However, during the cosmic history much higher energy scales can be reached, and thus off-shell dark matter production may be more efficient. In other words, through cosmological history, a part of the energy in the on-shell sector has been transferred to off-shell sector (while the reverse does not happen) and we detect this energy gravitationally as dark matter. The main purpose of this study to investigate the production of O$f$DM in the early universe and its observational consequences.  

In summary:
\begin{enumerate}

\item[$\bullet$] Whenever a scattering happens, there is a chance of producing dark matter particles which is given by \eqref{production_rate_simplified} and \eqref{production_rate_2}. Furthermore, the probability of producing two dark matter particles in one scattering is much lower than producing only one.

\item[$\bullet$] Dark matter production is much more efficient at high (center of mass) energy scatterings. Therefore, most of the dark matter is produced during the stages in the cosmological history  where the universe is dense (lots of scatterings) and hot (high energies), i.e. early universe.

\end{enumerate}

Before ending this section, let us discuss the physical range for the non-locality scale $\Lambda$. If $\Lambda$ comes from quantum gravitational effects or fundamental discreteness of spacetime \cite{Sorkin:2007qi,Aslanbeigi:2014zva,Benincasa:2010ac}, we expect it to be around Planck energy, $M_{\rm P}$. On the other hand, {\it a priori}, $\Lambda$ can be much smaller than $M_{\rm P}$, even as low as $\sim 10 $ TeV, as suggested in large extra dimension models that are constructed to address the hierarchy problem (e.g., \cite{Burgess:2004kd}), or by the cosmological non-constant problem \cite{Afshordi:2015iza}. However, in this paper we assume $\Lambda \gg H_{\rm inf}$, i.e. the non-locality scale is much larger than the Hubble scale during inflation. Otherwise, it would not be consistent to use the standard results of slow-roll inflation when  $\Lambda \lesssim H_{\rm inf}$, since the effect of non-locality on the evolution of inflaton or metric could not be neglected.

\section{Off-shell Dark Matter Production}\label{Dark Matter Production}

What are the processes in the early universe that are relevant for O$f$DM production? First of all, we consider inflation as a starting point in the universe. Whatever happened before inflation is diluted by the exponential expansion of the universe and is not relevant for our discussion. Furthermore, the effect of non-locality on the inflationary predictions can be neglected in the $H_{\rm inf} \ll \Lambda$ regime. After inflation, we consider two major processes that produce dark matter particles: inflaton decay to standard model particles (reheating) and radiation self interaction in the universe.

%
%

\subsection{Reheating}\label{sec::reheating}
In this section, we consider the simplest reheating model: inflaton ($\phi$ field)  decays through the effective interaction $g\phi \psi \bar \psi$, where $\psi$ represents standard model fields or an intermediate field\footnote{In this case we assume that the mass of $\psi$ field is much smaller than the inflaton's.} that decays into standard model particles later.

Decay of inflaton into (on-shell) standard model particles makes the radiation fluid of the universe, given that particle energies are much larger than their masses. As we mentioned earlier, however, inflaton will not only decay into on-shell particles; it also may decay into off-shell particles, or off-shell dark matter. Based on \eqref{production_rate_simplified}, decay rate into dark matter compared to the decay rate into radiation is suppressed by a factor of 
\beq\label{fraction_reheating}
f = \frac{1}{4\pi}\left(\frac{m_{\phi}}{\Lambda}\right)^2\ll1,
\eeq
where $m_{\phi}$ is the mass of inflaton at the end of inflation. As a result, after inflation there are three major constituents of the universe:
\begin{enumerate}
\item {\it Inflaton field ($\phi$):} This field can be treated as a non-relativistic matter after inflation when $m\gg H$ \cite{Kofman:1997yn}. Inflaton energy density ($\rho_{\phi}$) is the dominant energy density of the universe after inflation and it perturbatively decays into radiation (decay rate $\Gamma$) and dark matter (decay rate $f\Gamma$). We later comment on why the coherent decay of inflaton can be ignored.

\item {\it Radiation:} This includes all (on-shell) $\psi$ particles. Since the decay rate of inflaton into radiation is much bigger than the decay rate into dark matter, radiation energy density ($\rho_r$) will dominate the energy density of the universe after the decay of inflaton field.
\item {\it Dark matter:} This includes all off-shell $\psi$ particles. As we argue later, dark matter acts as a non-relativistic matter and its energy density is the last one to become dominant. 
\end{enumerate}
This system of three fluids satisfies the following equations:
\bea
&&\dot \rho_{\phi}+3H\rho_{\phi}=-(1+f)\Gamma \rho_{\phi},\label{infdecay}\\
&&\dot \rho_r+4H\rho_r=\Gamma \rho_{\phi},\label{inftorad}\\
&&\dot \rho_{\phi\to DM}+3H\rho_{\phi \to DM}=f\Gamma \rho_{\phi},\label{inftoDM}
\eea
which can be solved along with the Friedmann equation, where $H=\frac{\dot a}{a}$ is the Hubble parameter, $a$ is the scale factor of the universe and $\rho_{\phi \to DM}$ is the contribution to dark matter energy density from inflaton decay.\footnote{Annihilation of radiation into O$f$DM barely changes the radiation energy density, which is why it has been ignored in \eqref{inftorad}.}

Let us define the fraction of total dark matter energy density from inflaton decay
\beq
x=\frac{\rho_{\phi\to DM}}{\rho_{DM}},
\eeq
where $\rho_{DM}$ is the total dark matter energy density. Solving the system of differential equations, we arrive at \cite{Erickcek:2011us}

\beq\label{reheating}
T_{rh}=x\frac{T_{eq}}{f},
\eeq
where $T_{rh}$ is the reheating temperature (temperature of radiation at the time of inflaton-radiation equality) and $T_{eq}$ is the temperature at the matter-radiation equality.

\begin{figure}
	\begin{subfigure}[t]{0.47\textwidth}
        		\includegraphics[width=\hsize]{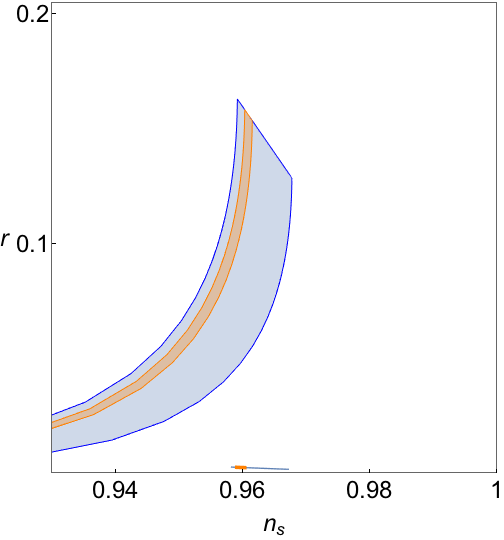}
		\centering
		\caption{Blue upper (lower) region shows the prediction of natural ($R^2$) inflation for $k=0.002~{\rm Mpc}^{-1}$ with $T_{rh}=10$~MeV-$10^{15}$GeV. Orange regions show the prediction of the same models with the constraint coming from O$f$DM model for $\Lambda=0.1M_{\rm P}-M_{\rm P}$.} 
\label{natural}	
    	\end{subfigure} %
	\hfill
	\begin{subfigure}[t]{0.47\textwidth}
        		\includegraphics[width=\hsize]{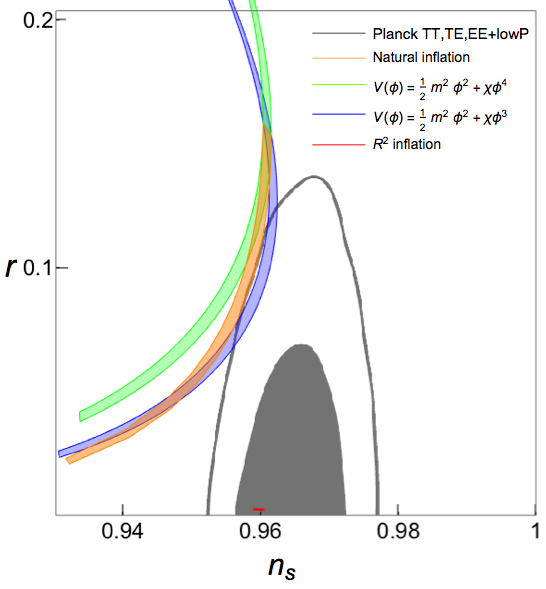}
		\caption{Prediction of $n_s$ and $r$ for different inflationary potentials at $k=0.002 ~{\rm Mpc}^{-1}$. Each region represents the prediction with the assumption of O$f$DM with $\Lambda=0.1M_{\rm P}- M_{\rm P}$.  The shaded region (curve) show the 68\% (95\%) constraints from CMB observations \cite{Ade:2015lrj}.}

\label{ns_r}
    	\end{subfigure} %
	\caption{Predictions of spectral index, $n_s$, and tensor to scalar ratio, $r$, for a number of inflationary potentials with O$f$DM constraint \eqref{reheating}.}
\end{figure}


Since $T_{eq} \simeq$ 0.75 eV, Equation \eqref{reheating} fixes the reheating temperature for a given mass of inflaton and $x \approx 1$.\footnote{We will show later that $x$ is very close to 1.} This can be used, for example, to constrain spectral index, $n_s$, and tensor to scalar ratio, $r$, of a given inflationary potential by using the following equation:
\bea
N_e=67&-&\ln\left(\frac{k}{a_0H_0}\right)+\frac{1}{4}\ln\left(\frac{V}{M_{\rm P}^4}\right)+\frac{1}{4}\ln\left(\frac{V}{V_e}\right)\notag\\
&+&\frac{1}{12}\ln\left(\frac{\rho_{th}}{V_e}\right)-\frac{1}{12}\ln g_{th}
\eea
where $N_e$ is the number of e-foldings that mode $k$ is superhorizon during inflation, $V_e$ is the potential energy at the end of inflation, $\rho_{th} \sim g_{th} T^4_{rh}$ is the radiation energy density at reheating temperature, $a_0H_0$ is the present Hubble radius, $V$ is the potential energy when mode $k$ crosses the horizon during inflation, $g_{th}$ is the number of effective bosonic degrees of freedom at reheating temperature and we have assumed pressureless effective equation of state for inflaton during reheating \cite{Planck:2013jfk}.

 Figure \ref{natural} shows how the predicted regions for the Natural \cite{Freese:1990rb} and $R^2$ \cite{Starobinsky:1980te} inflations have shrunk significantly in the $(n_s,r)$ plane  as a result of fixing the reheating temperature. A similar constraint can be found for other inflationary potentials, e.g. Figure \ref{ns_r} shows the prediction of O$f$DM model for a number of inflationary models. 


We shall next review and justify the assumptions we made in the above calculations.

\subsubsection{Coherent decay of inflaton}

The coherent decay of inflaton is negligible if the following condition is satisfied \cite{Erickcek:2011us,Kofman:1997yn}

\beq\label{preheating_condition}
\frac{\Gamma}{m_{\phi}}\ll\left(\frac{m_{\phi}}{M_{\rm P}}\right)^2.
\eeq
Using $\Gamma \sim \frac{T_{rh}^2}{M_P}$ and \eqref{reheating}, this reduces to
\beq
10^{-18}\left(\frac{\Lambda}{M_p}\right)^4\left(\frac{10^{-5}M_P}{m_{\phi}}\right)^7\ll1,
\eeq
which is generically satisfied for models of large field inflation with $m_\phi \sim 10^{-5} M_P$.

\subsubsection{Non-relativistic dark matter}\label{NR_DM}
The mass distribution of dark matter particles is given in \eqref{mass_density}. When a dark matter particle is produced, its energy is below $E_{\rm CM}$, while, according to \eqref{mass_density}, masses of the 98\% of the dark matter particles are above $0.1E_{\rm CM}$. In other words, upon production, most dark matter particles are mildly relativistic, but through the expansion of the universe they soon become non-relativistic. This justifies our earlier assumption to model dark matter particles as a non-relativistic fluid. 

\subsection{Radiation self-interaction}

How much dark matter is produced as a result of radiation self interaction? Here we find an upper bound on the amount of dark matter production through self interaction of radiation. Let us assume a simple annihilation process, such as in Figure \ref{scattering}, and ignore the intrinsic mass of the particles. Ignoring the intrinsic mass of the particles is consistent with finding an upper limit for the dark matter production, since we are allowing for more dark matter production by ignoring the intrinsic masses (more phase space volume to produce O$f$DM). The average mass of the produced dark matter particles is
\beq\label{average_mass}
\int dm~mP_{1F}(m)=\frac{8}{15}E_{\rm CM},
\eeq
and the cross section of producing one dark matter particle is\footnote{This is again consistent with finding the upper bound, since the cross section of two off-shell production is much smaller. }
\beq\label{1Fscattering}
\sigma_{1F}=\frac{\sigma_O}{4\pi}\left(\frac{E_{\rm CM}}{\Lambda}\right)^2=\frac{\lambda^2}{128\pi^2\Lambda^2}.
\eeq
Since this contribution to dark matter has been produced at very high energies (lower bound on reheating temperature is $T_{rh}>5$~MeV), it will be highly redshifted today. As a result, current energy density of dark matter is the same as its mass density (see Section \ref{NR_DM}). The comoving mass density of the produced dark matter particles through radiation self interaction is given by
\beq
\frac{d\rho_{rad\to DM}}{dt}=a^{3}(t)\int \frac{d^3p_1}{(2\pi)^3}\frac{d^3p_2}{(2\pi)^3}g_1n(\vec p_1)g_2n(\vec p_2) \langle m\sigma_{1F}v_{rel}\rangle,
\label{comoving_mass1}
\eeq
where $t$ is the cosmological time, $n(\vec p)=\frac{1}{e^{|\vec p|/T}\pm1}$ is the occupation number of incoming on-shell states at temperature $T$, $g$ is the degeneracy factor, $v_{rel}$ is the relative velocity of the incoming particles and $\vec p_i$'s are the momenta of the incoming particles. It is clear that \eqref{comoving_mass1} results in a bigger comoving mass density when we choose bosonic occupation number. 

Using \eqref{average_mass}-\eqref{1Fscattering}, $v_{rel}\lesssim2$ and performing the integrals over momenta in \eqref{comoving_mass1}, we arrive at
\beq\label{comoving_mass2}
\frac{d\rho_{rad\to DM}}{dt} \lesssim g_1g_2\frac{8\lambda^2}{45(2\pi)^6}\Gamma^2[3.5]\zeta^2[3.5]a^3(t)\frac{T^7}{\Lambda^2},
\eeq
where $\Gamma$ and $\zeta$ are gamma and Riemann zeta functions, respectively. 

Perturbative calculations are valid only if $\lambda<1$. If we consider this condition in \eqref{comoving_mass2} and sum over all constituent of the radiation fluid, we arrive at
\beq\label{comoving_mass}
\rho_{rad\to DM} < 4\times10^{-5}\int dt~g^2 a^3(t)\frac{T^7}{\Lambda^2},
\eeq
where $g$ is the total number of degrees of freedom in the radiation fluid. 

During reheating (by solving \ref{infdecay}-\ref{inftoDM})
\beq
t \propto a^{3/2},~~ T^4\propto \rho_{rad}\propto a^{-3/2}.
\eeq
Substituting these values back in \eqref{comoving_mass}, we realize that the annihilation of radiation into dark matter is most efficient at the end of reheating.
The same manipulation shows that the annihilation of radiation into dark matter during radiation era happens at the beginning of radiation era and is of the same order.  

Let us now work out how much dark matter will be produced in radiation era (after reheating). During radiation era
\beq
t=\sqrt{\frac{45}{2\pi^2g}}\frac{M_{\rm P}}{T^2}.
\eeq
Combining this, with Eq. (\ref{comoving_mass}), and the results of Sec. (\ref{sec::reheating}), we find:
\bea
\frac{\rho_{rad\to DM}}{\rho_{DM}} &<& 10^{-5}\times  \frac{g^{3/2} M_{\rm P} T_{rh}^2}{ T_{eq} \Lambda^2} \nonumber\\ 
&\sim& 10^{-3}\times  \frac{g^{3/2} M_{\rm P} T_{eq}  \Lambda^2}{m^4_\phi} \nonumber\\
&\sim&  10^{-7} \left(g \over 124\right)^{3/2}   \left(\Lambda \over M_{\rm P} \right)^{2}  \left(m_\phi \over 10^{-5} M_{\rm P} \right)^{-4},  
\eea
where we used $g \simeq 124$ for standard model of particle physics.

%

Therefore,  for $\Lambda \sim M_{\rm P}$ and high scale inflation  $m_{\phi}\approx 10^{-5}M_{\rm P}$, the production of O$f$DM due to radiation self-interaction is much smaller than the contribution from inflaton decay (in effect $x=1$). However, $\rho_{rad\to DM}$ can become important in scenarios with lighter inflaton, i.e. if $m_\phi \lesssim 10^{-7} (M_{\rm P} \Lambda)^{1/2}$.

So far we have studied the predictions of this model in the context of inflation. As we showed earlier, this model effectively fixes the reheating temperature of the universe. By constraining the reheating temperature, we can narrow the predictions of $(n_s,r)$ for a given inflationary potential, by fixing the number of e-foldings. However, the predictions for $(n_s,r)$ are model dependent and vary with the inflationary potential. Conversely, one can use the observational constraints on $(n_s,r)$ as a way to fix the non-locality scale $\Lambda$, in the context of a given inflationary model.

\section{Cold O$f$DM}\label{Cold OFDM}
In principle, O$f$DM particles with very low masses can be produced in scatterings. These low mass particles can behave like hot dark matter at different stages in the evolution of the universe. Let us estimate an upper bound on the fraction of hot O$f$DM particles at a given redshift.

An off-shell dark matter particle with mass $m$ has energy $E_m=\frac{E_{\rm CM}^2+m^2}{2E_{\rm CM}}$ and momentum $p_m=\frac{E_{\rm CM}^2-m^2}{2E_{\rm CM}}$ , where $E_{\rm CM}$ is the energy of the process producing the dark matter particle.\footnote{This comes from conservation of energy-momentum in the rest frame of incoming particle(s). Here, we have ignored mass of the on-shell particle produced together with O$f$DM particle.} At redshift $z$, this particles is relativistic if $p_m\frac{1+z}{1+z_{pr}} \gtrsim m$, where $z_{pr}$ is the redshift at the time of production.

\begin{figure}
	\begin{subfigure}[t]{0.46\textwidth}
        		\includegraphics[width=\hsize]{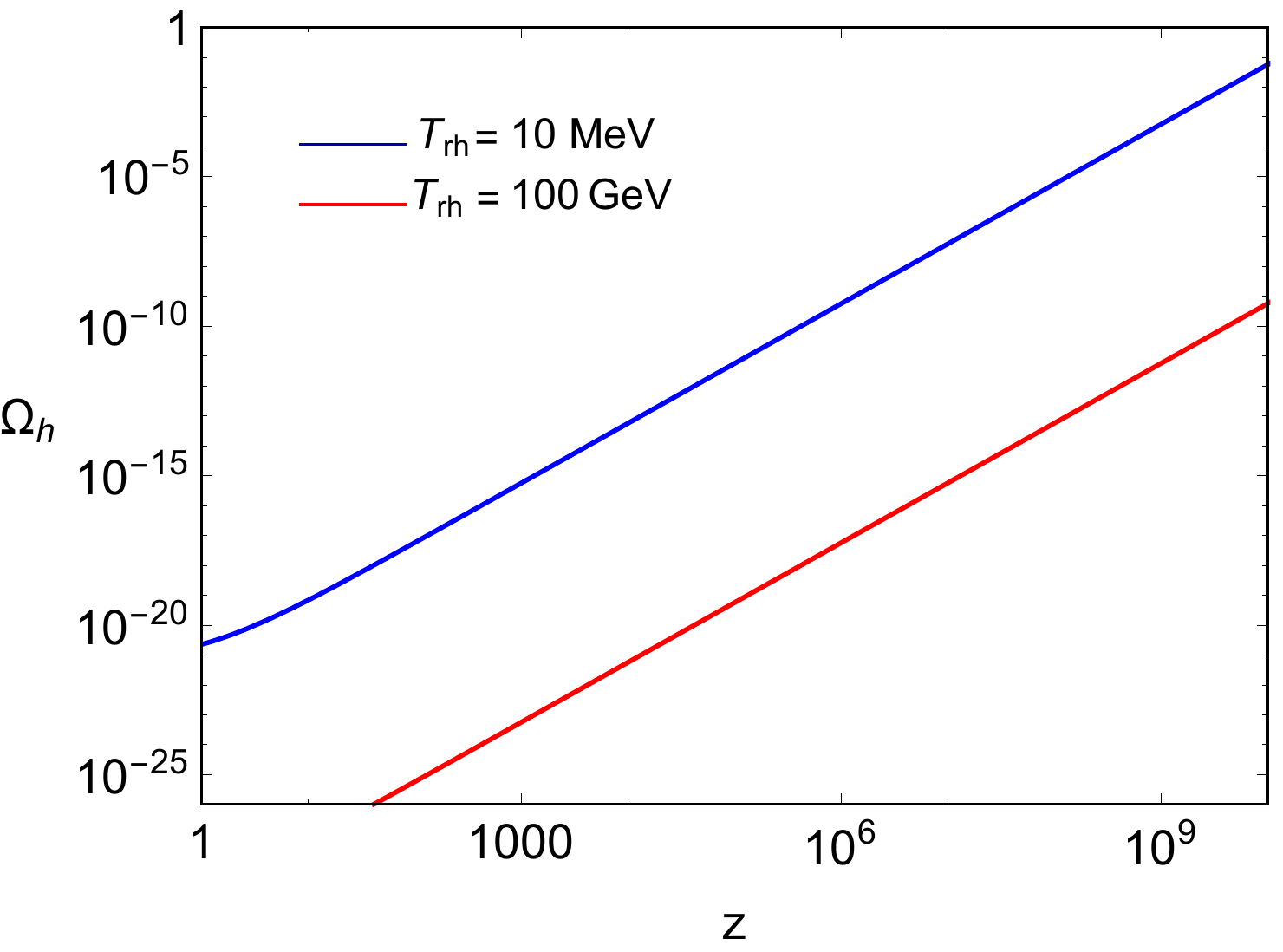}
		\caption{The fraction of off-shell dark matter particles, produced at the time of reheating, that remain relativistic down to a given redshift.}
\label{hot_dark_matter}
    	\end{subfigure} %
	\hfill
	\begin{subfigure}[t]{0.52\textwidth}
        		\includegraphics[width=\hsize]{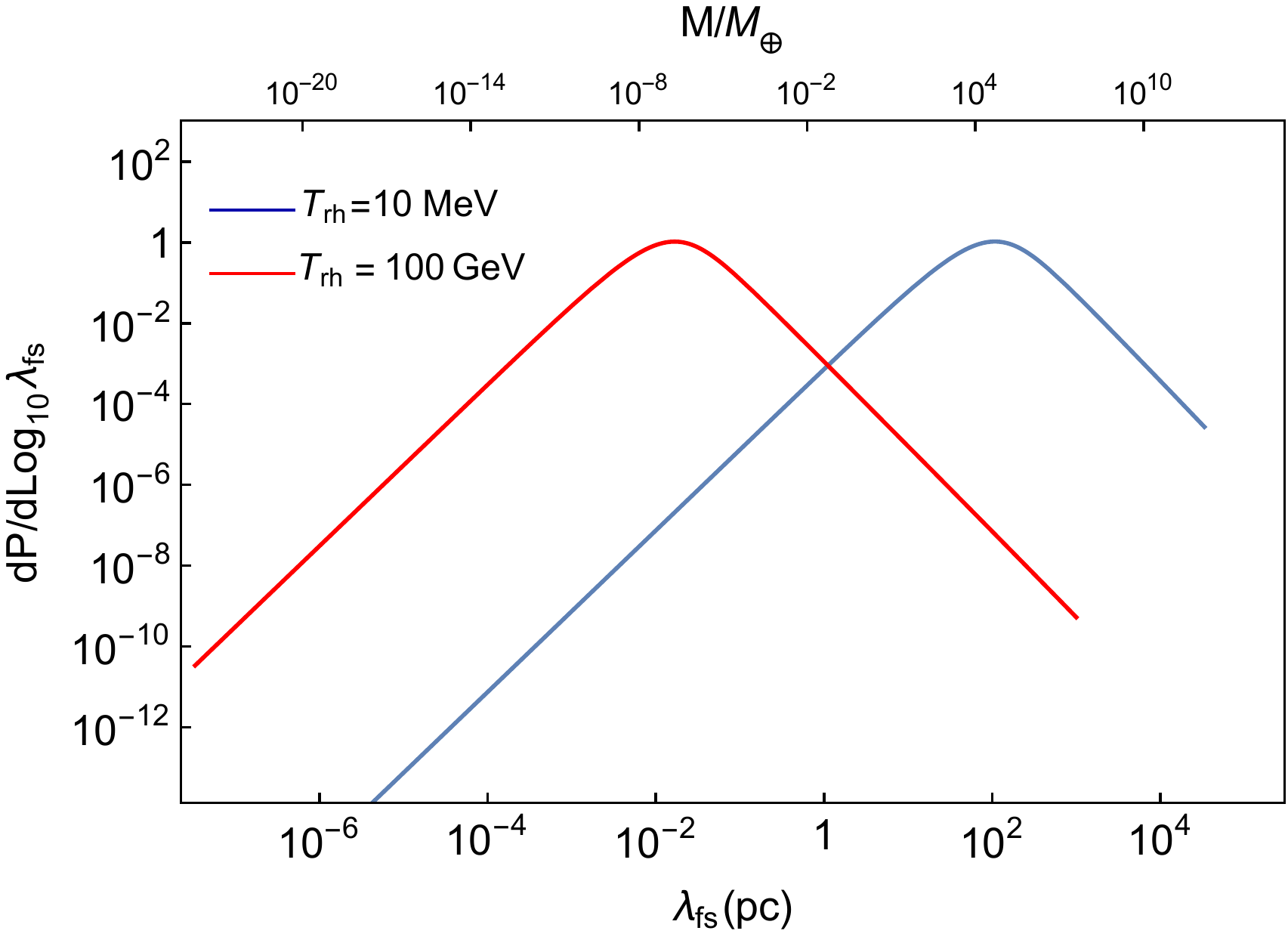}
\caption{Distribution of free streaming distance of O$f$DM for different reheating temperatures. The top axis shows the characteristic halo mass associated with the free streaming scale, in units of Earth mass.}
\label{free_stream_distribution}
    	\end{subfigure} %
	\caption{}
\end{figure}

Given the mass distribution of O$f$DM particles and assuming that most of the dark matter particles are produced at the time of reheating (as we discussed in previous sections), we can find the fraction of hot dark matter particles ($\Omega_h$), which is shown in Figure \ref{hot_dark_matter}. Only a small fraction of O$f$DM is hot at $z< 1000$, which makes it a good candidate for CDM. This result is not surprising since, as we mentioned earlier, even at the time of production these particles are {\it only mildly} relativistic.

\begin{figure}
\includegraphics[width=0.9\hsize]{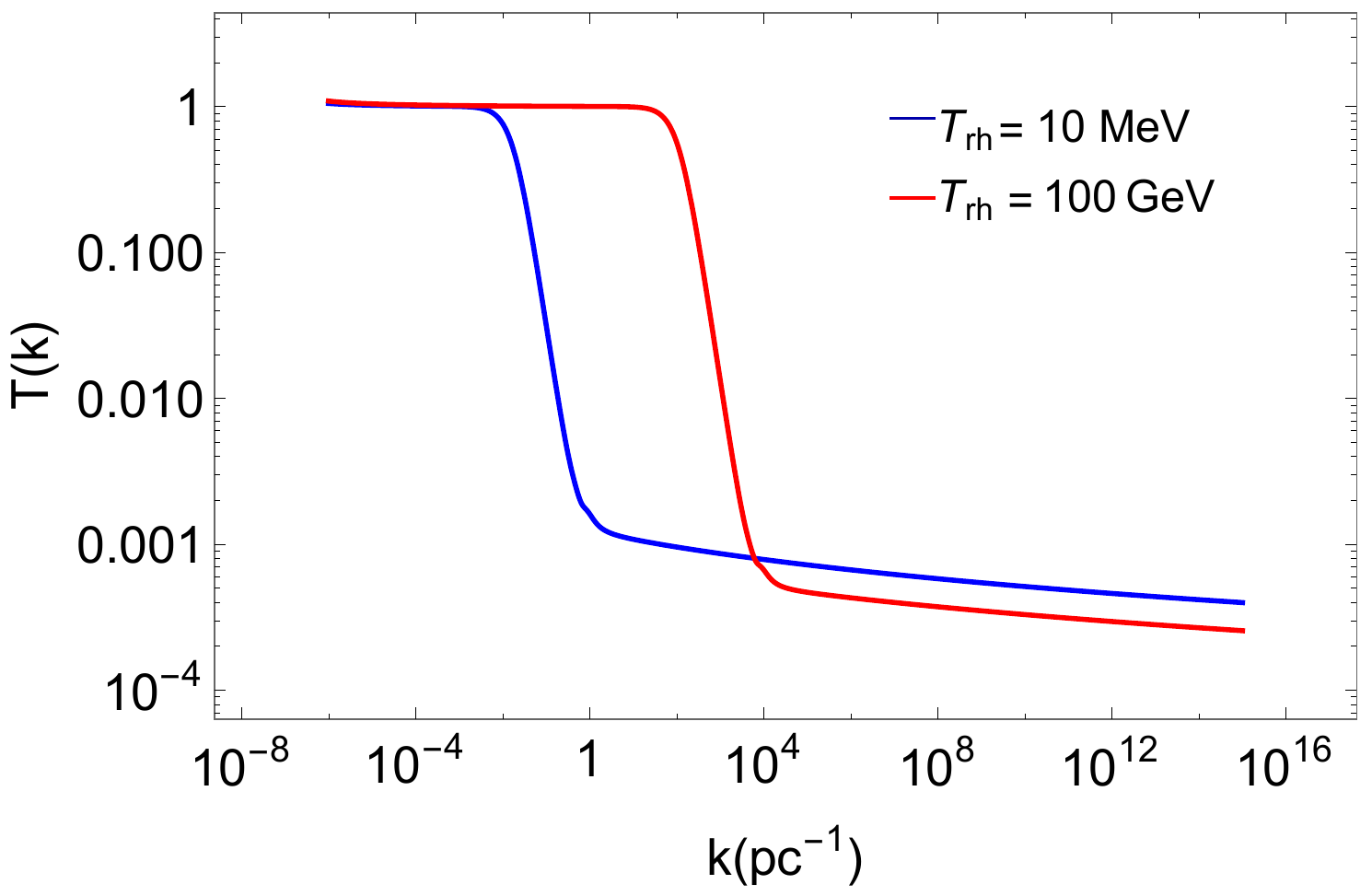}
\caption{Matter transfer function due to the growth in early matter era and free streaming effect. Instead of an exponential cut-off for large $k$ in thermal scenarios, there is $\propto (\ln k)^{-1}$ drop in O$f$DM scenario.}
\label{Transfer_function}
\end{figure}

Let us work out the distribution of free streaming distance $\lambda_{fs}$. This is given by
\beq\label{free_stream}
\lambda_{fs}=u\int\frac{dt}{\sqrt{a^4+u^2 a^2}}
\eeq 
where $u=a_{pr}\frac{v}{\sqrt{1-v^2}}$ and $v=\frac{p_m}{E_m}$ is the velocity of dark matter particle with mass $m$ at the time of production. Assuming $a_{pr}=a_{rh}$, Equation \eqref{free_stream} gives the free streaming distance in terms of $m$ and $T_{rh}$. This equation can be used further to derive the probability distribution of $\lambda_{fs}$, since the probability distribution of $m$ \eqref{mass_density} is known. The result is shown in Figure \ref{free_stream_distribution}. 
Since the velocity distribution of O$f$DM particles is different from Maxwell-Boltzmann distribution, probability distribution of $\lambda_{fs}$  in this model is different from ordinary thermal WIMP scenario. In particular, it has a much shallower power-law (rather than gaussian) cut-off at large $\lambda_{fs}$'s. This leads to a different matter power spectrum (on small-scales) which can, in principle, be a way to distinguish these two models. Figure \ref{Transfer_function} shows the matter transfer function $T(k)$.

 In Figure \ref{Transfer_function} two effects has been considered: Growth in matter fluctuations due to an early era of matter domination (inflaton dominated era) and free streaming effect.
Early matter era result into amplification of matter fluctuations for modes that enter the horizon during reheating. This amplification is roughly $\propto \frac{k^2}{\ln(k)}$ \cite{Erickcek:2011us}. On the other hand, free streaming effect result into the decrease in the matter power spectrum on small scales $\propto k^{-2}$. The combination of the two effects is seen in Figure \ref{Transfer_function}. On small scales, transfer function drops as $(\ln k)^{-1}$ which is to be contrasted with a much steeper gaussian cut-off in thermal scenarios.

Future gravitational probes of dark matter structure on small scales can potentially test this prediction for matter power spectrum on $10^{-1}-10^{-3}$ pc scales  \cite{Erickcek:2010fc,Baghram:2011is,Rahvar:2013xya}.

\section{Summary and Future Prospect}\label{conclusion}
In this paper, we laid out the phenomenological implications of the off-shell dark matter (O$f$DM) model. This model is motivated by considering the effect of Planck scale nonlocality on the evolution of quantum fields which manifests itself by introducing a new set of excitations. The new excitations, dubbed off-shell modes, cannot be detected through scattering experiments, making them a natural candidate for dark matter. So, if O$f$DM makes up the majority of the observed cosmological dark matter, we would not be able to detect dark matter particles directly.

However, O$f$DM particles can be produced in scattering experiments and this is one way to indirectly confirm their existence by detecting missing energy in scatterings. The probability of missing energy is given by \eqref{production_rate_simplified} and \eqref{production_rate_2}. High energy collider experiments with enough  sensitivity to detect this missing energy could be  a possible way to test this model, albeit not the most practical one.

We also discussed predictions of O$f$DM model in the context of cosmology and showed that it is intertwined with the physics of inflation and reheating. For a very simple reheating model, we showed that O$f$DM particles are generically produced in the era of reheating and through the decay of inflaton. Since O$f$DM particles do not interact with other particles (or each other), they do not reach a thermal distribution. We calculated O$f$DM distribution function in our simple reheating model and  showed that it leads to much shallower suppression of matter power spectrum on small scales compared to a gaussian cutoff of thermal dark matter candidates. This, in principle, could be another way to test the model via the observations probing matter power spectrum in sub-pc scales. 

We end this paper by noting the following theoretical aspects of O$f$DM which are yet to be explored: 
\begin{enumerate}
\item Throughout this paper we assumed that off-shell modes of a nonlocal field gravitate like ordinary (on-shell) matter, i.e. an off-shell mode with mass $m$  gravitates  like a normal particle with the same mass. This assumption, which seems reasonable, is yet to be verified through a consistent coupling of nonlocal quantum field theories to gravity.
\item So far, the quantization of this type of nonlocal field theory has only been done only scalars. But how about spinor or gauge fields?
This is especially important in the case of gauge theories which govern all interactions in the standard model of particle physics. There are (at least) two obvious ways to proceed here: 
\begin{enumerate}
\item One can define a nonlocal version of gauge transformations to keep gauge invariance. This presumably implies that scattering processes have to include pairs of on-shell modes, or otherwise charge conservation would be violated. In the case of our phenomenological reheating model in Section \ref{sec::reheating}, it means that the inflaton field has to first decay into a neutral field which later decays into standard model particles, otherwise Equation \eqref{fraction_reheating} is not applicable. 

\item Gauge invariance is broken at a Planck suppressed level, similarly to the violation of diffeomorphism invariance in Horava-Lifhsitz gravity \cite{Horava:2009uw}.  In this case, one should look for (possibly dangerous) physical consequences of breaking gauge invariance.
\end{enumerate}
\item Off-shell modes of a nonlocal field cannot be detected in realistic collider experiments. But how about other types of experiments? Scatterings are just a subset of experiments that can be done in labs. Is there a way of observing off-shell modes in laboratory directly?

\end{enumerate}

%
%
%
%

\acknowledgments
This research is supported by Perimeter Institute for Theoretical Physics. Research at Perimeter Institute is supported by the Government of Canada through
Industry Canada and by the Province of Ontario through the Ministry of Research and
Innovation.

\bibliography{Dark_Matter}
\bibliographystyle{JHEP}

\end{document}